\documentstyle[aps,pre,multicol,epsf]{revtex}
\begin{document}
\newcommand{\tbox}[1]{\mbox{\tiny #1}}
\newcommand{\half}{\mbox{\small $\frac{1}{2}$}}
\newcommand{\mbf}[1]{{\mathbf #1}}

\title{Parametric dependent Hamiltonians, wavefunctions, \\ 
random-matrix-theory, and quantal-classical correspondence}

\author
{
Doron Cohen$^1$ and Tsampikos Kottos$^2$\\
\footnotesize
$^1$
Department of Physics, Harvard University, Cambridge, MA 02138 \\
$^2$
Max-Planck-Institut f\"ur Str\"omungsforschung,
37073 G\"ottingen, Germany
}

\date{December 1999, revised September 2000}

\maketitle


\begin{abstract}

We study a classically chaotic system which is described 
by a Hamiltonian ${\cal H}(Q,P;x)$ where $(Q,P)$ are 
the canonical coordinates of a particle in a 2D well, 
and $x$ is a parameter. By changing $x$ we can deform 
the `shape' of the well. The quantum-eigenstates of the 
system are $|n(x)\rangle$. We analyze numerically how the 
parametric kernel $P(n|m)= |\langle n(x)|m(x_0)\rangle|^2$ 
evolves as a function of $\delta x \equiv (x{-}x_0)$. 
This kernel, regarded as 
a function of $n-m$, characterizes the shape of the wavefunctions, 
and it also can be interpreted as the local density of states (LDOS). 
The kernel $P(n|m)$ has a well defined classical limit, 
and the study addresses the issue of 
quantum-classical correspondence (QCC).  
Both the perturbative and the non-perturbative regimes 
are explored. The limitations of the random-matrix-theory (RMT) 
approach are demonstrated. 

\end{abstract}

\begin{multicols}{2}

\section{Introduction}

Consider a system whose total Hamiltonian is ${\cal H}(Q,P;x)$, 
where $(Q,P)$ is a set of canonical coordinates, 
and $x$ is a constant parameter. This parameter 
may represent the effect of some externally controlled field. 
We assume that both ${\cal H}_0 = {\cal H}_0(Q,P;x_0)$ 
and ${\cal H} = {\cal H}(Q,P;x)$ 
generate classically chaotic dynamics of similar nature. 
Moreover, we assume that $\delta x \equiv (x{-}x_0)$ 
is {\em classically small}, 
meaning that it is possible to apply linear analysis 
in order to describe how the energy surfaces ${\cal H}(Q,P;x)=E$ 
are deformed as a result of changing the value of $x$. 
Quantum mechanically, we can use a basis where 
${\cal H}_0 = \mbf{E}_0$ has a diagonal representation, 
while 
\begin{eqnarray} \label{e1} 
{\cal H} \ \  = \ \ \mbf{E}_0 \ + \ \delta x \ \mbf{B}
\end{eqnarray}
For reasonably small $\hbar$, it follows from general 
semiclassical considerations \cite{mario},  
that $\mbf{B}$ is a {\em banded matrix}. 
Generically, this matrix {\em looks random}, 
as if its off-diagonal elements were {\em independent} random numbers.

It was the idea of Wigner \cite{wigner} forty years ago,  
to study a simplified model, where the Hamiltonian is 
given by Eq.(\ref{e1}), and where 
$\mbf{B}$ is a {\em random} banded matrix. This is 
known as Wigner's banded random matrix (WBRM) model.     
The applicability of such a model is a matter 
of {\em conjecture}. Obviously this conjecture 
should be tested$^\dag$. The most direct way to 
test it, which we are going to apply, is to 
take the matrix $\mbf{B}$ of a `physical' Hamiltonian, 
and then to randomize the signs of its off-diagonal 
elements. The outcome of such operation will be referred to  
as the {\em effective} WBRM model that is associated 
with the {\em physical} Hamiltonian. One issue of this paper 
is to make a comparison between the eigenstates of 
the physical Hamiltonian, and those of the associated  
effective WBRM model.

The {\em standard} WBRM model (unlike the `effective' one) 
involves an additional simplification. Namely, one 
assumes that $\mbf{B}$ has a rectangular band profile. 
The theory of eigenstates for the standard WBRM model 
is well known \cite{wigner,casati,fyodo}. 
Increasing $x$, starting from $\delta x=0$
the eigenstates of Eq.(\ref{e1}) change their nature. 
The general questions to address are: 
\begin{enumerate}
\item 
What are the {\em parametric regimes} in the 
parametric evolution of the eigenstates;  
\item
How the structure of the eigenstates 
changes as we go via the subsequent regimes. 
\end{enumerate}
Recently some ideas have been introduced \cite{frc,vrn,wls} 
how to go beyond Wigner's theory in case of physical Hamiltonians. 
It has been suggested that there are at least three generic 
parametric scales   
$\delta x_c^{\tbox{qm}}\ll
\delta x_{\tbox{prt}}\ll
\delta x_{\tbox{SC}}$ 
that control the parametric evolution of the eigenstates. 
We shall define these parametric scales later. 
Accordingly one should distinguish between 
the standard perturbative regime ($\delta x \ll \delta x_c^{\tbox{qm}}$), 
the core-tail regime 
($\delta x_c^{\tbox{qm}} \ll \delta x \ll \delta x_{\tbox{prt}}$), 
and the semiclassical regime  
($\delta x \gg \delta x_{\tbox{SC}}$).

The purpose of this paper is not just to numerically 
establish (for the first time) the existence of the 
parametric regimes suggested in \cite{frc,vrn,wls}, 
but mainly to address question {\bf (2)} above \cite{prm}. 
Namely, we would like to study how the structure of the 
eigenstates changes as we go via the subsequent regimes. 
In particular we would like to understand the significance
of RMT assumptions in the general theoretical considerations. 
The latter issue has been left unexplored in the 
`quantum chaos' literature. (Note however that literally 
the same question is addressed in numerous publication once 
spectral statistics of eigenvalues, 
rather than eigenstate structure, is concerned).   
We also suggest a new procedure for `region analysis' of the 
eigenstate structure. We are going to distinguish between 
first-order tail regions (FOTRs), higher-order far-tail 
regions, and non-perturbative (core) region. Our main 
conclusion is going to be that RMT is inadequate for 
the analysis of any features that go beyond 
first-order perturbation theory.

\section{The model Hamiltonian}

We study the Hamiltonian 
\begin{eqnarray} \label{e2} 
{\cal H}(Q,P;x) = \half(P_1^2{+}P_2^2 + Q_1^2{+}Q_2^2)  
+ x \cdot Q_1^2 Q_2^2
\end{eqnarray}
with $x=x_0+\delta x$ and $x_0=1$. 
This Hamiltonian describes the motion of 
a particle in a 2D well (see Fig.1). 
The units are chosen such that the mass is equal to one, 
the frequency for small oscillations is one, 
and for $\delta x=0$ the coefficient of the 
anharmonic term is also one. The energy $E$ 
is the only dimensionless parameter of the 
classical motion. Our numerical study is 
focused on an energy window around $E \sim 3$ 
where the motion is mainly chaotic.

In the classical analysis there is only one 
parametric scale, which is $\delta x_c^{\tbox{cl}} \sim 1$. 
This scale determines the regime of 
(classical) linear analysis. 
For $\delta x \ll \delta x_c^{\tbox{cl}}$ 
the deformation of the energy surface  
${\cal H}_0(Q,P;x)=E$ can be described 
as a linear process. Later we are going to 
give a precise mathematical formulation of 
this idea. From now on assume that we 
are in the classical linear regime.

\ \\
\epsfysize=1.3in
\epsffile{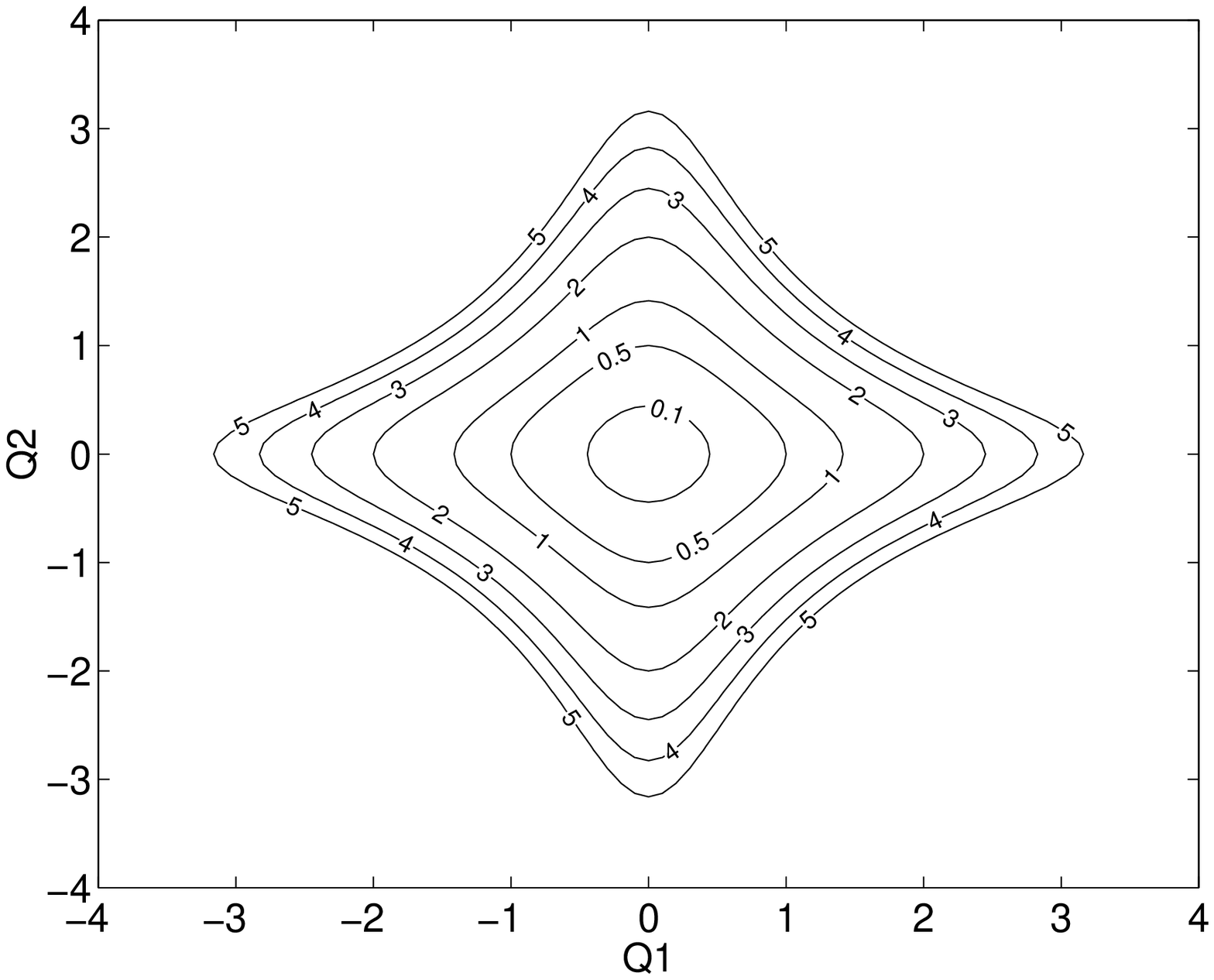} 
\ \ \ 
\epsfysize=1.3in
\epsffile{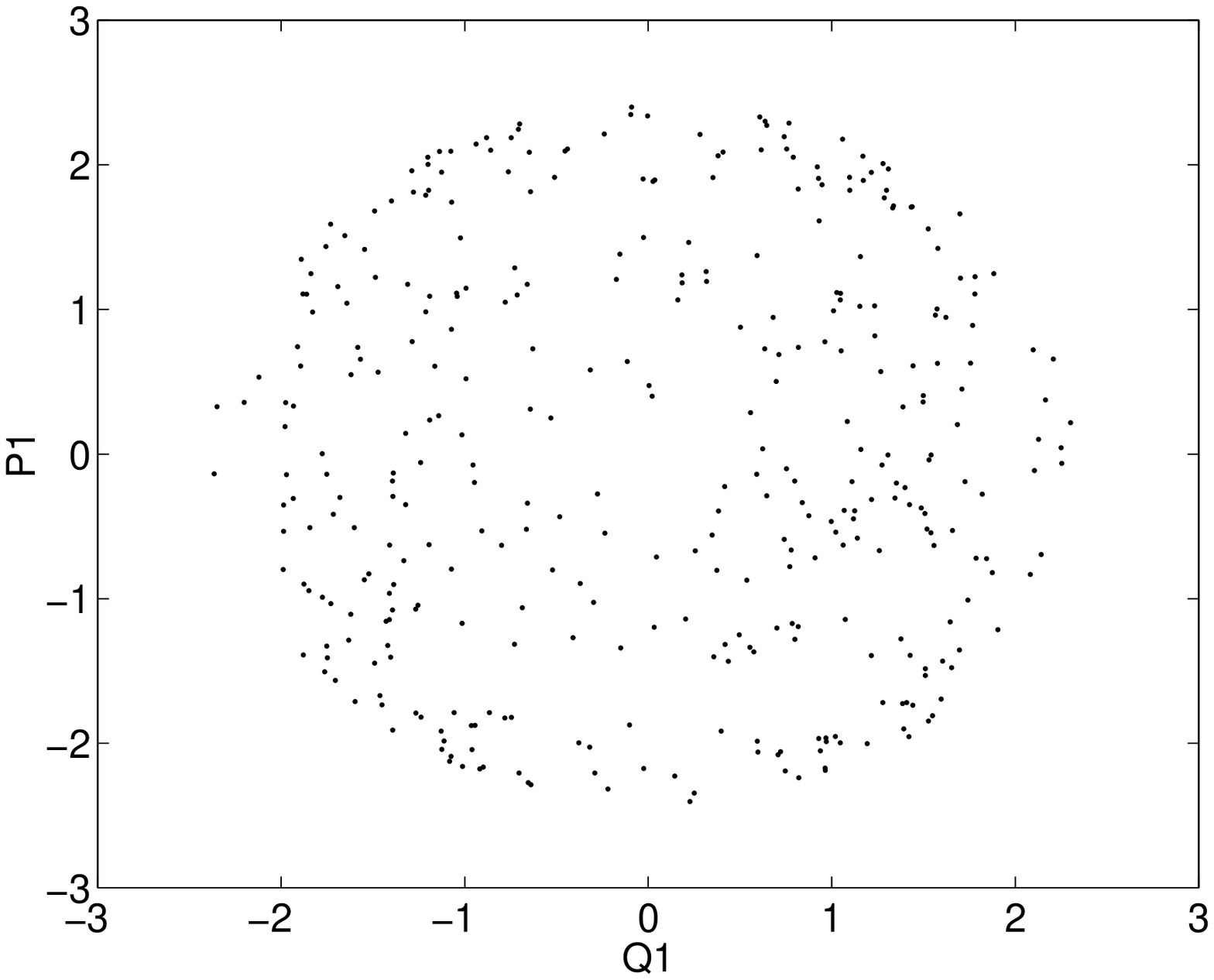}
\noindent \\
{\footnotesize {\bf FIG.1}:
{\em Left:} equipotential contours of the model 
Hamiltonian (\ref{e2}) with $x=x_0=1$.  
{\em Right:} A Poincare section of a long trajectory 
($0<t<1300$) that we have picked in order to get the 
fluctuating quantity ${\cal F}(t)$. 
The initial conditions are \mbox{$(Q_1,Q_2,P_1,P_2)=(1,0,1,2)$} 
corresponding to $E=3$. The trajectory is quite ergodic.  
It avoids some small quasi-integrable islands 
(the main one is around $(0,0)$). } \\

Let us pick a very long ergodic trajectory 
$(Q(t),P(t))$ that covers densely the energy 
surface $E$. See Fig.1. 
Let us define the fluctuating quantity 
\begin{eqnarray} \label{e_2} 
{\cal F}(t) \ \equiv \ 
-(\partial {\cal H}/\partial x)
\ = \ -Q_1^2 Q_2^2
\end{eqnarray} 
For the later analysis it is important to know 
the distribution of the variable ${\cal F}$, 
and to characterize its temporal correlations. 
The average value is $F = \langle {\cal F} \rangle$. 
The angular brackets stand for microcanonical average 
over $(Q(0),P(0))$, which should be the same as 
time ($t$) average (due to the assumed ergodicity).   
The auto-correlation function of ${\cal F}(t)$ is
\begin{eqnarray} \label{e_3} 
C(\tau) \ \ = \ \ 
\langle \ ({\cal F}(t)-F) \ ({\cal F}(t{+}\tau)-F) \ \rangle
\end{eqnarray} 
Note that $C(\tau)$ is independent of $t$, and 
that average over $t$ should give the same result 
as a microcanonical average over $(Q(0),P(0))$.

The variance of the fluctuations 
is $C(0) = \langle ({\cal F}-F)^2 \rangle$. 
The correlation time will be denoted by 
$\tau_{\tbox{cl}}$. Note that with our choice 
of units $\tau_{\tbox{cl}} \sim 1.0$  
within the energy range of interest. 
The power spectrum $\tilde{C}(\omega)$
of the fluctuating ${\cal F}(t)$, is obtained  
via a Fourier transform of $C(\tau)$. See Fig.2.
The average $F$ and the variance $C(0)$ determine 
just the first two moments of the ${\cal F}$ 
distribution. The probability density of ${\cal F}$
will be denoted by $P_{\tbox{F}}({\cal F})$.

All the required information for the 
subsequent semiclassical analysis is contained 
in the functions $C(\tau)$ and $P_{\tbox{F}}({\cal F})$ 
as defined above. All we have to do in 
order to numerically determine them is 
to generate one very long ergodic trajectory (see Fig.1),  
to compute the respective ${\cal F}(t)$, 
and from it to extract the desired information 
(see Fig.2 and Fig.3).     
It is convenient to express $P_{\tbox{F}}({\cal F})$
in terms of a scaling function as follows
\begin{eqnarray} \label{e_4} 
P_{\tbox{F}}({\cal F}) \ = \ \frac{1}{\sqrt{C(0)}} 
\hat{P}_{\tbox{cl}}\left(-\frac{{\cal F}-F}{\sqrt{C(0)}}\right)
\end{eqnarray} 
By this definition the scaled distribution  
$\hat{P}_{\tbox{cl}}(f)$ is characterized 
by a zero average ($\langle f \rangle = 0$), 
a unit variance ($\langle f^2 \rangle = 1$), 
and it is properly normalized. Note that 
$\hat{P}_{\tbox{cl}}(-f)$  rather than 
$\hat{P}_{\tbox{cl}}(f)$ correspond 
to $P_{\tbox{F}}({\cal F})$. This has been done 
for later convenience.

\section{The quantized Hamiltonian}

Upon quantization we have a second 
dimensionless parameter $\hbar$.  
For obvious reasons we are considering
a de-symmetrized ($1/8$) well with Dirichlet 
boundary conditions on the lines
$Q_1{=}0$ and $Q_2{=}0$ and $Q_1{=}Q_2$. 
The matrix representation of ${\cal H}={\cal H}(Q,P;x)$ 
in the basis which is determined 
by ${\cal H}(Q,P;0)$ is very simple.
The eigenstates ($n=1,2,3\cdots$) of the chaotic 
Hamiltonian ${\cal H}_0={\cal H}(Q,P;1)$ 
has been found numerically.

The phase space volume ($dQdP$ integral) 
which is enclosed by an energy surface 
${\cal H}(Q,P;x)=E$  is given by a function 
$n=\Omega(E,x)$. It is convenient to measure 
phase space volume in units of $(2\pi\hbar)^d$, 
where $d=2$ is the dimensionality of our system. 
Upon quantization the phase space volume $n$ 
corresponds to the level index ($n=1,2,3\cdots$). 
This is known as Weyl law. It follows that 
$g(E)=\partial_E\Omega(E,x)$ corresponds  
to the density of states, and 
$\Delta = 1/g(E) \propto \hbar^d $ 
is the mean level spacing.

In the following presentation we are 
going to assume the our interest is restricted 
to an energy window which is `classically small' 
but `quantum mechanically large'.  
In the numerical analysis of our model Hamiltonian 
the energy window was $2.8<E<3.1$, where the 
classical motion is predominantly chaotic.   
The mean level spacing for $E\sim 3$ 
is given approximately by the formula 
$\Delta\approx 4.3*\hbar^2$. 
Our numerical analysis has been carried  
out for $\hbar=0.03$ and for $\hbar=0.015$. 
Smaller values of $\hbar$ where beyond our 
numerical capabilities since the maximal 
matrix that we can handle is of size $5000 \times 5000$.

The representation of $Q_1^2 Q_2^2$, 
in the basis which is determined by the chaotic 
Hamiltonian ${\cal H}_0$, gives 
the matrix $\mbf{B}$ of Eq.(\ref{e1}). 
The banded matrix $\mbf{B}$ and the band profile 
are illustrated in Fig.2.
The band profile is implied by the semiclassical 
relation \cite{mario}:  
\begin{eqnarray} \label{e_5}  
|\mbf{B}_{nm}|^2 \ \ \approx \ \ 
\frac{\Delta}{2\pi\hbar} 
\tilde{C}\left(\frac{E_n-E_m}{\hbar}\right)  
\end{eqnarray} 
As we see from Fig.2 the agreement  
with this formula is remarkable.
For the bandwidth Eq.(\ref{e_5}) implies that 
$\Delta_b=2\pi\hbar/\tau_{\tbox{cl}}$. 
It is common to define $b=\Delta_b/\Delta$. 

\ \\
\epsfysize=1.65in
\epsffile{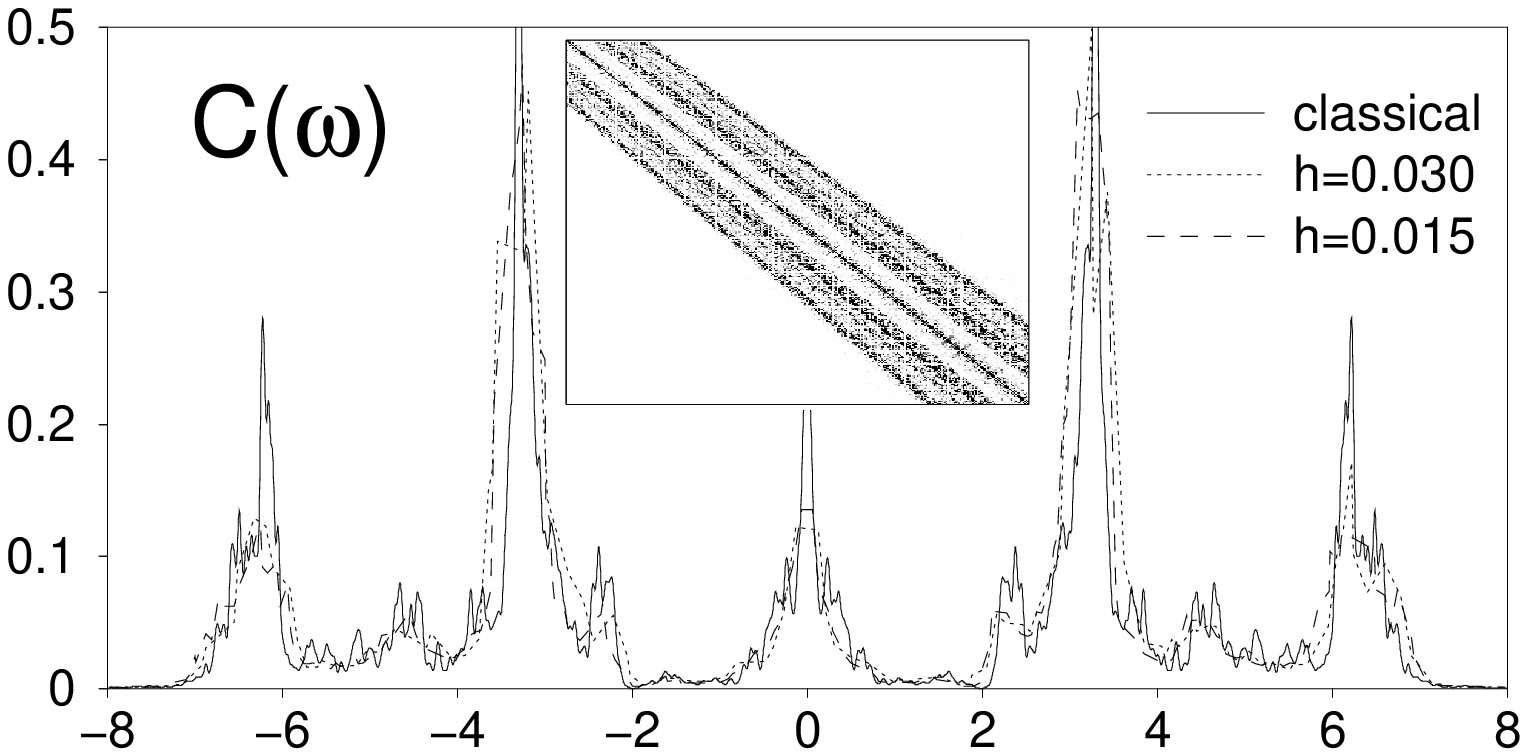}
\noindent \\
{\footnotesize {\bf FIG.2}:
The band profile $(2\pi\hbar/\Delta)\cdot|\mbf{B}_{nm}|^2$ 
versus $\omega = (E_n{-}E_m)/\hbar$ is compared 
with the classical power spectrum $C(\omega)$. 
{\em Inset}: An image of a piece of the $\mbf{B}$ matrix. } \\

\section{Definition of the LDOS profile}

The quantum-eigenstates of the Hamiltonian 
${\cal H}(Q,P;x)$ are $|n(x)\rangle$, 
and the ordered eigen-energies are $E_n(x)$.  
We are interested in the parametric kernel 
\begin{eqnarray} \label{e3} 
P(n|m) \ = \ |\langle n(x)|m(x_0)\rangle|^2 
\ = \ \mbox{trace}(\rho_n\rho_m)
\end{eqnarray}
In the equation above $\rho_m(Q,P)$ and $\rho_n(Q,P)$ 
are the Wigner functions that correspond to the 
eigenstates $|m(x_0)\rangle$ and $|n(x)\rangle$ 
respectively. The trace stands for $dQdP/(2\pi\hbar)^d$ 
integration.

We can identify $P(n|m)$ as the local density of states (LDOS), 
by regarding it as a function of $n$, where $m$ is considered 
to be a fixed reference state. An average of $P((m+r)|m)$ over 
several $m$-states leads to the LDOS profile $P(r)$. 
Alternatively, fixing $n$, the vector $P(n|m)$ describes 
the shape of the $n$-th eigenstate in the ${\cal H}_0$ representation. 
By averaging $P(n|(n-r))$ over few eigenstates one obtains 
the average shape of the eigenstate (ASOE).  
The ASOE is just $P(-r)$. Thus the ASOE and the LDOS are 
given by the same function. One would have to be more careful 
with these definitions if ${\cal H}_0$ were  
integrable while ${\cal H}$ non-integrable.

The kernel $P(n|m)$ gives the overlap between 
the $n$th eigenstate of ${\cal H}$ and the 
$m$th eigenstate of ${\cal H}_0$.  
For $\delta x=0$ we have simply $P(n|m)=\delta_{nm}$. 
For $\delta x > 0$ the kernel develops a structure, 
which is described by the LDOS profile $P(r)$. 
If $\delta x$ is very small then evidently 
$P(r)$ consists of Kronecker delta (at $r=0$) 
and tail regions ($|r|>0$). Later we are going to 
distinguish between first-order tail regions (FOTRs), 
and higher order far-tail regions. 
As $\delta x$ becomes larger a non-perturbative
core region appears around $r=0$.  
Namely, the profile exhibits a bunch of states 
(rather than one) that share most of the probability. 
If $\delta x$ becomes even larger, the distinction between 
core and tail regions become meaningless, and 
the LDOS profile becomes purely non-perturbative. 
We are going to explain that the non-perturbative 
profile reflects the underlying classical 
phase space structure.

\section{The classical approximation for the LDOS}

The classical approximation \cite{felix2,frc,vrn,wls} 
for $P(n|m)$ follows naturally from the definition Eq.(\ref{e3}).  
It is obtained if we approximate $\rho_n(Q,P)$ 
by a microcanonical distribution that 
is supported by the energy surface 
${\cal H}(Q,P;x)=E_n(x)$. Namely, 
\begin{eqnarray} \nonumber 
\rho_n(Q,P) \ &=& \ 
\frac{1}{g(E)}\delta({\cal H}(Q,P;x)-E_n(x)) 
\\  \label{e_mc} 
\ &=& \ \delta(\Omega({\cal H}(Q,P;x)) - n)
\end{eqnarray}
and a similar expression (with $x=x_0$) for 
$\rho_m(Q,P)$. In the classical limit $n$ is 
the phase space volume by which we label energy surfaces.
Each energy surface $n$ is associated 
with a microcanonical state $\rho_n(Q,P)$. 
The classical LDOS profile will be denoted 
by $P_{\tbox{cl}}(r)$. The $\delta x$ regime where 
the classical approximation $P(r) \approx P_{\tbox{cl}}(r)$
applies will be discussed in a later section.

By definition, for $\delta x \ll \delta x_c^{\tbox{cl}}$ 
the deformed energy surfaces departs linearly 
from the $\delta x=0$ surfaces. As already 
stated in the Introduction, being in this classical 
linear regime is a fixed assumption of this 
paper. Now we want to explain the consequences 
of this assumption. One may consider these consequences 
as giving an operational definition for the classical 
linear regime.     
The dispersion (square-root of the variance) of 
the classical profile in the classical linear regime is 
\begin{eqnarray} \label{e_6} 
\delta E_{\tbox{cl}} \ \ = \ \ \sqrt{C(0)} \times \delta x  
\end{eqnarray}
(This should be divided by $\Delta$ if we want the 
dispersion in proper $r$ units. See (\ref{e_7}) below).
For our model Hamiltonian, for energies $E \sim 3$, 
we have found that $\delta E_{\tbox{cl}}\approx 0.38*\delta x$.
Eq.(\ref{e_6}) can be regarded as a special consequence 
of the following scaling relation which we are going to derive below:
\begin{eqnarray} \label{e4} 
P_{\tbox{cl}}(r) \ = \ 
\frac{\Delta}{\sqrt{C(0)}\ \delta x} \cdot 
\hat{P}_{\tbox{cl}}\left(\frac{\Delta \cdot r}{\sqrt{C(0)}\ \delta x}\right) 
\end{eqnarray}
The scaling function has already been defined 
in Eq.(\ref{e_4}), and it is illustrated in Fig.3.
The classical profile $P_{\tbox{cl}}(r)$ is in general 
non-symmetric, but it follows from Eq.(\ref{e4}) 
that it must be characterized by $\langle r \rangle = 0$. 
[By definition the scaling function of Eq.(\ref{e_4}) 
gives zero average]. Another obvious feature 
is having sharp cutoffs, beyond which $P_{\tbox{cl}}(r)=0$. 
The existence of these outer `classically forbidden' regions 
follows from the observation that for large enough $r$ 
there is no longer classical overlap between the 
energy surfaces that correspond to $|m(x_0)\rangle$ 
and $|n(x)\rangle$ respectively.

The rest of this section is dedicated to technical 
clarifications of Eq.(\ref{e4}), and it can be skipped 
in first reading. The derivation is done in two steps. 
The first step is to establish a relation 
between $P_{\tbox{cl}}(r)$ and its trivially related 
version $P_{\tbox{E}}(\epsilon)$. The second step is to  
demonstrate that $P_{\tbox{E}}(\epsilon)$ is related 
to $P_{\tbox{F}}({\cal F})$ of Eq.(\ref{e_4}).  
It is also possible to make a one-step derivation 
that relates $P_{\tbox{cl}}(r)$ to $P_{\tbox{F}}({\cal F})$, 
but we find the derivation below more physically 
appealing.

\ \\
\epsfysize=1.3in
\epsffile{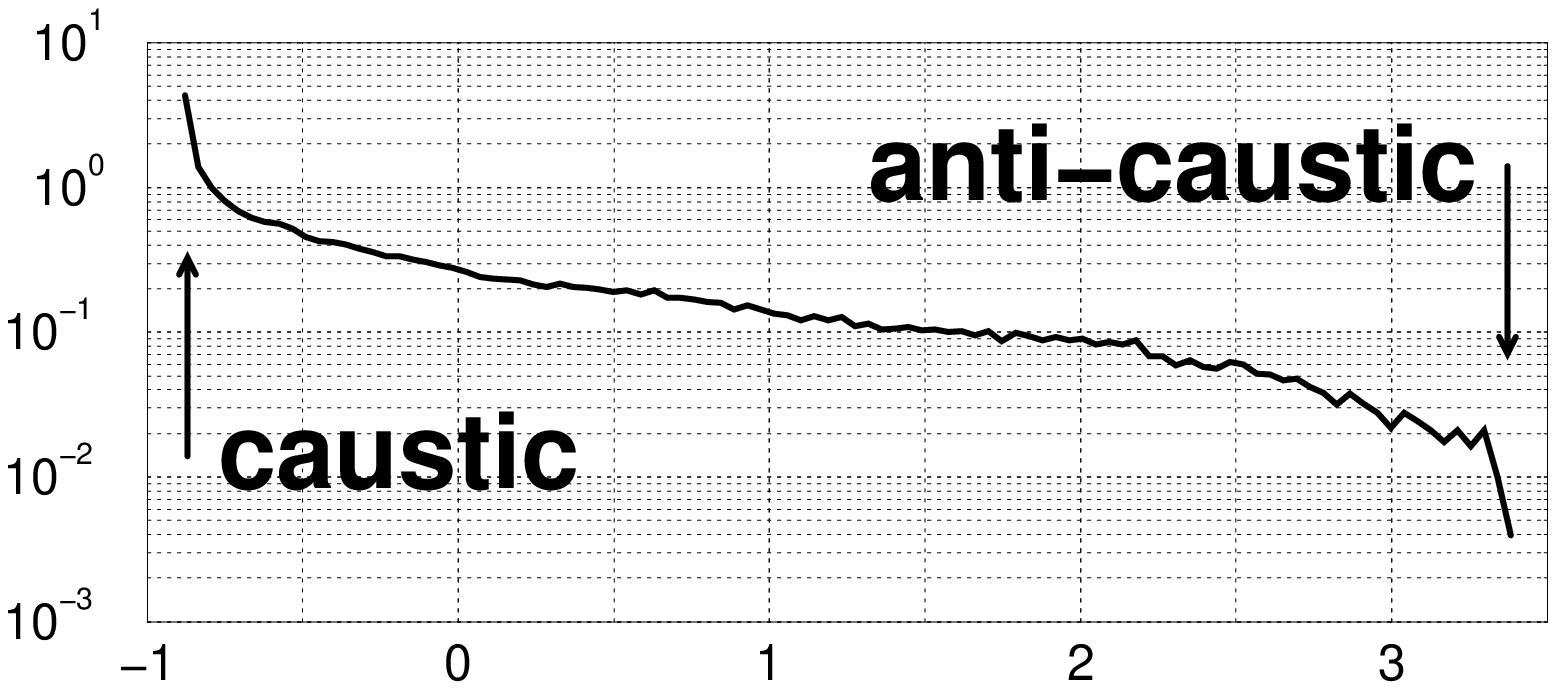} \\
\noindent
{\footnotesize {\bf FIG.3}:
The scaled classical profile $\hat{P}_{\tbox{cl}}()$. 
One unit on the horizontal axis correspond to 
energy difference $\delta E_{\tbox{cl}}\approx 0.38*\delta x$.
Note that $r = 0$ implies $(E_n(x){-}E_m(x_0))>0$. 
The caustic is located at $(E_n(x){-}E_m(x_0))=0$, 
while the anti-caustic is located at $(E_n(x){-}E_m(x_0)) = 1.65*x$.   
The ``forbidden regions'' are defined 
as those regions where $P_{\tbox{cl}}(r) = 0$.
They are located to the left of the caustic 
and to the right of the anti-caustic. } \\

By differentiating of $n=\Omega(E,x)$, 
keeping $n$ constant, we get the relation 
$\delta E = - F(x) \delta x$, where 
$F(x) = \partial_x\Omega(E,x)/g(E)$ 
is known as the (generalized) conservative force.
Using the latter expression it is a straightforward 
exercise to prove that 
$F(x) = \langle {\cal F} \rangle \equiv F$.  
Alternatively, we can eliminate $E$ from 
the relation $n=\Omega(E,x)$, and write the 
result as $E=E_n(x)$. Accordingly  
$F(x) = -(\partial E_n(x)/\partial x)$. 
Now we can write the following relation:
\begin{eqnarray} \nonumber
E_n(x) - E_m(x_0) \ = \ 
\left. \frac{\partial E}{\partial x} \right|_n \delta x
+ \left. \frac{\partial E}{\partial n} \right|_x (n-m)
\end{eqnarray}
which can be re-written in the following form
\begin{eqnarray} \label{e_7}
\epsilon \ \  = \ \  
-F(x) \ \delta x \ \ + \ \ ({1}/{g(E)}) \ r
\end{eqnarray}
Whenever we regard the kernel $P(n|m)$ as a function 
of $n-m$ we use the notation $P(r)$. 
But sometimes it is convenient to regard $P(n|m)$ 
as an energy distribution $P_{\tbox{E}}(\epsilon)$.  
Due to the change of variables (\ref{e_7}) we have 
the following relation:
\begin{eqnarray} \label{e_8}
P(r) \ \ = \ \ 
\frac{1}{g(E)} \ P_{\tbox{E}}\left( 
\frac{1}{g(E)} r - F(x)\delta x \right)
\end{eqnarray}
The energy distribution $P_{\tbox{E}}(\epsilon)$ 
can be formally defined as follows:
\begin{eqnarray} \label{e_9}
P_{\tbox{E}}(\epsilon) \ = \ 
\sum_n P(n|m) \ \delta(\epsilon-(E_n(x){-}E_m(x_0)))
\end{eqnarray}
In the classical limit the summation over 
$n$ should be interpreted as a $dn$ integral.
For $P(n|m)$ in the above expression 
we can substitute the definition Eq.(\ref{e3}) 
with $\rho_n$ and $\rho_m$ approximated 
as in Eq.(\ref{e_mc}). A straightforward manipulation 
leads to the result:
\begin{eqnarray} \nonumber
P_{\tbox{E}}(\epsilon) \ &=& \ 
\langle \delta(\epsilon - 
({\cal H}(Q,P;x)-{\cal H}(Q,P;x_0))) \rangle
\\  \nonumber
\ & = & \
\langle \delta(\epsilon + \delta x {\cal F}(t) ) \rangle
\  =  \ 
\frac{1}{\delta x} 
P_{\tbox{F}}\left(-\frac{1}{\delta x}\epsilon\right)
\end{eqnarray}
Together with (\ref{e_4}) and (\ref{e_8}), 
we get Eq.(\ref{e4}) along with the implied 
special result (\ref{e_6}).

\section{Numerical determination of LDOS profiles}

Given $\delta x$ we can determine numerically 
the LDOS profile $P(r)$. Representative profiles 
are displayed in Fig.4.  For the purpose 
of further discussion we introduce  the following definitions:
\begin{itemize}
\item
The classical LDOS profile $P_{\tbox{cl}}(r)$   
\item
The quantum mechanical LDOS profile $P(r)$
\item
The effective WBRM LDOS profile $P_{\tbox{RMT}}(r)$
\item
The first-order perturbative profile $P_{\tbox{prt}}(r)$ 
\end{itemize} 
We have already discussed the classical LDOS profile. 
Below we explain how we numerically determine the 
quantum mechanical LDOS profiles $P(r)$ and $P_{\tbox{RMT}}(r)$,  
and we also define the profile $P_{\tbox{prt}}(r)$.

The numerical procedure for finding $P(r)$ 
is straightforward. For a given $\delta x$ we  
have to diagonalize the matrix~(\ref{e1}).  
The columns of the diagonalization matrix $\mbf{T}_{mn}$ are 
the eigenstates of the Hamiltonian, and 
by definition we have $P(n|m)= |\mbf{T}_{mn}|^2$. 
Then $P(r)$ is computed by averaging over 
roughly 300 reference states that are located 
within the classically-small energy window $2.8<E<3.1$.    
Fig.4  displays typical profiles.

The effective WBRM Hamiltonian is obtained by randomizing 
the signs of the off-diagonal elements in the $\mbf{B}$ matrix. 
For the effective WBRM Hamiltonian exactly the same 
procedure (as for $P(r)$) is applied leading to $P_{\tbox{RMT}}(r)$.

In order to analyze the structure of either $P(r)$ or 
$P_{\tbox{RMT}}(r)$ we have defined the first-order 
perturbative profile as follows:
\begin{eqnarray} \label{e5} 
P_{\tbox{prt}}(r) \ = \  
\frac{\delta x^2 \ |\mbf{B}_{nm}|^2}{\Gamma^2 + (E_n{-}E_m)^2}
\end{eqnarray}
It is implicit in this definition that 
$(E_n{-}E_m)$ and $|\mbf{B}_{nm}|^2$ 
should be regarded as a function of $r$.  
The $r=0$ value of the band-profile should be  
re-defined by an interpolation. 
The parameter $\Gamma \equiv b_0\Delta$ is determined 
(for a given $\delta x$) such that the $P_{\tbox{prt}}(r)$ 
has a unit normalization. Note that Wigner's Lorentzian 
would be obtained if the band profile were flat.

\section{Region analysis for the quantal LDOS}

By comparing $P(r)$ to $P_{\tbox{prt}}(r)$ as in Fig.4, 
we can determine$^\ddag$ 
the range $b_1\mbox{[left]} < r < b_1\mbox{[right]}$ 
where $P_{\tbox{prt}}(r)$ is a reasonable approximation for $P(r)$. 
Loosely speaking (avoiding the distinction between 
the `left' and the `right' sides of the profile) we shall say 
that $P_{\tbox{prt}}(r)$ is a reasonable approximation for $|r| < b_1$. 
The core is defined as the region $|r|<b_0$. 
The FOTRs are $b_0<|r|<b_1$. The far-tail regions are $|r|>b_1$.

\ \\
\epsfysize=5.8in
\epsffile{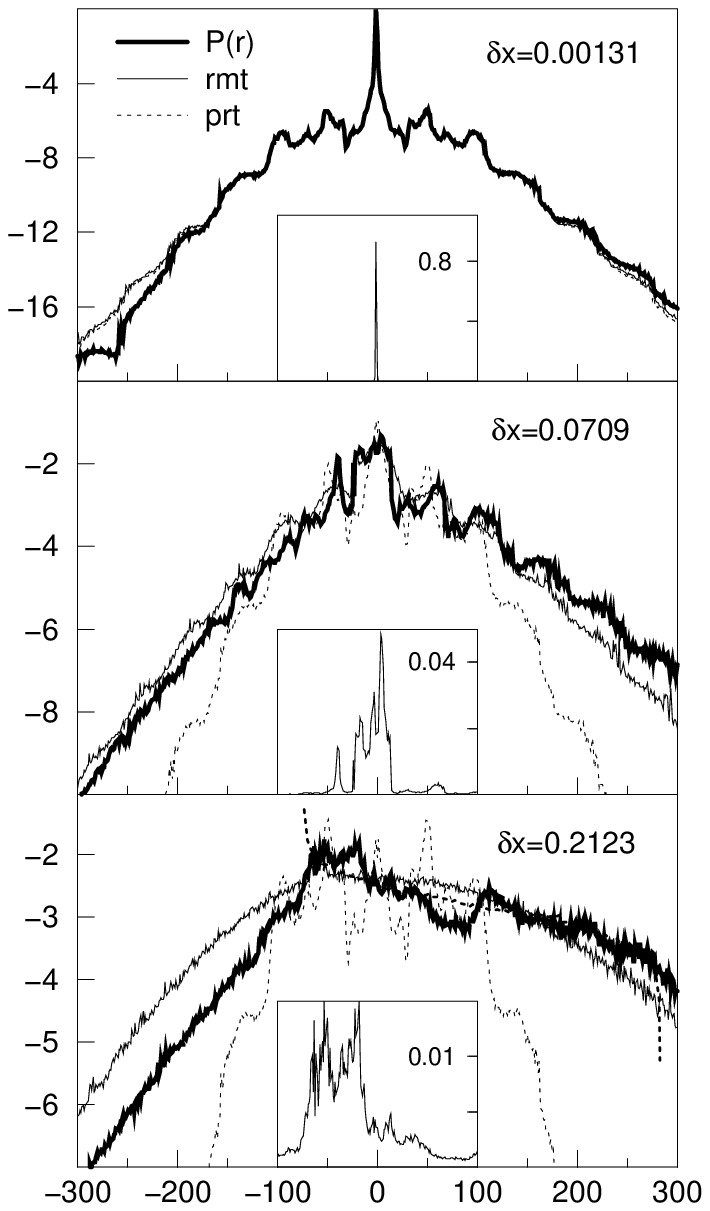} \\
\noindent
{\footnotesize {\bf FIG.4}:  
The quantal profile $P(r)$ is compared with 
$P_{\tbox{prt}}(r)$ and with $P_{\tbox{RMT}}(r)$. 
We are using here the $\hbar=0.015$ output. 
The insets are normal plots while 
the main figures are semilog plots.
In the lower plot ($\delta x = 0.2123$) 
the classical LDOS profile $P_{\tbox{cl}}(r)$ 
is represented by heavy dashed line. } \\

\ \\

The results of this region analysis are summarized by Fig.5.
In the following sections we are going to present a detailed 
discussion of this analysis.   
For the convenience of the reader we summarize: 
\begin{itemize}
\item
$b_0 \ = \ $ border of the core region
\item
$b_1 \ = \ $ border of the first order tail region (FOTR)
\end{itemize}
Having $b_0 \ll 1$ implies a standard perturbative 
structure. Having $1 \ll b_0 \ll b_1$ implies that 
we have a well developed core-tail structure.  
Having $b_0 \sim b_1$ implies a purely 
non-perturbative structure. In the latter case 
the distinction between core and tail regions 
become meaningless.

\section{The standard perturbative regime}

The standard perturbative regime  
$\delta x \ll \delta x_c^{\tbox{qm}}$ 
is defined by the requirement $b_0(\delta x) \ll 1$.
This condition implies that $P(n|m) \sim \delta_{nm}$. 
For numerical purpose it is convenient 
to define $\delta x_c^{\tbox{qm}}$ as the 
value of $\delta x$ for which $P(r=0) \approx 0.5$.
The theoretical considerations of \cite{frc} 
imply that $\delta x_c^{\tbox{qm}} \propto \hbar^{(1{+}d)/2}$. 
The prefactor is a classical quantity whose 
precise value depends on the operational 
definition of $\delta x_c^{\tbox{qm}}$. 
With the operational definition given above 
we have extracted the result  
$\delta x_c^{\tbox{qm}} \approx 3.8*\hbar^{3/2}$.

In the standard perturbative regime we can write  
schematically 
\begin{eqnarray}
P(n|m) \ \ \approx \ \ \delta_{nm} + \mbox{Tail}
\end{eqnarray}
The `Tail' is composed of FOTRs and far-tail 
regions. The former are given by Eq.(\ref{e5}), 
while the latter are determined by higher orders 
of perturbation theory. Note that for the {\em standard} 
WBRM we have by construction $b_1 \equiv b$, and more generally  
$n$-th order perturbation theory becomes essential 
for $(n{-}1) \times b < |r| < n \times b$. 
In case of our physical Hamiltonian, as well as 
for the associated {\em effective} WBRM model, the 
boundary $b_1$ is $\delta x$ dependent.

By comparing $P(r)$ with $P_{\tbox{RMT}}(r)$ we can see 
that RMT cannot be trusted for the analysis of the far-tails, 
because system-specific interference phenomena becomes 
important there. Namely, the RMT profile $P_{\tbox{RMT}}(r)$  
is almost indistinguishable from $P_{\tbox{prt}}(r)$. 
In contrast to that, the far-tails of $P(r)$ are dominated 
by either destructive interference (left tail), 
or by constructive interference (right tail).

\section{The core-tail regime}

The core-tail regime  
$\delta x_c^{\tbox{qm}} \ll \delta x \ll \delta x_{\tbox{prt}}$ 
is defined by the requirement $1 \ll  b_0 \ll b_1$. 
The theoretical considerations of \cite{frc} 
imply that $\delta x_{\tbox{prt}} \propto \hbar$. 
The prefactor is a classical quantity whose 
precise value depends on the operational 
definition of $\delta x_{\tbox{prt}}$. 
In our numerical analysis we have defined $\delta x_{\tbox{prt}}$ 
as the $\delta x$ for which the contribution of the FOTRs 
to the variance becomes less than $80\%$. 
With this operational definition 
we have extracted (using the lower subplot of Fig.5) 
the result $\delta x_{\tbox{prt}} \approx 5.3*\hbar$.

In the core-tail regime we can write schematically
\begin{eqnarray}
P(n|m) \ \ \approx \ \ \mbox{Core} + \mbox{Tail}
\end{eqnarray}
Disregarding the far-tail regions, the large-scale
behavior of $P(r)$ can be approximated by that 
of $P_{\tbox{prt}}(r)$.  As in the standard 
perturbative regime one observes that the far-tails are 
dominated by either destructive interference (left tail), 
or by constructive interference (right tail).

The core is a non-perturbative region. It means, 
that unlike the far-tail, it cannot be obtained 
from any finite-order perturbation theory. 
Once the core appears, the validity of first-order 
perturbation theory becomes a non-trivial matter.  
In \cite{frc} a non-rigorous argument 
is suggested in order to support the claim 
that, disregarding smoothing effect, 
the local mixing of neighboring levels does not 
affect the growth of the tail. 
An important ingredient in this argumentation is 
the (self consistent) assumption that  
most of the probability is well-contained in 
the core region. Indeed the analysis which is 
presented in Fig.5 is in agreement with this 
assumption.

The observation that the local mixing of 
neighboring levels does not affect the growth 
of the tail, implies that the tail grows 
as $\delta x^2$ and not like say $\delta x$. 
(The latter type of dependence is implied 
by an over-simplified argumentation). 
Having indeed $\delta x^2$ behavior is implied by 
observing that $P(r) \approx P_{\tbox{prt}}(r)$ 
for the FOTRs.

Finally, it should be emphasized that the local mixing of levels on 
the small scale $b_0$ is not reflected by Eq.(\ref{e5}). 
In particular, one should not expect Eq.(\ref{e5})  
to be literally valid within the core region ($|r|<b_0$).

\section{The non-perturbative regime }

In the non-perturbative regime  
($\delta x \gg \delta x_{\tbox{prt}}$) 
one may say that the core spills over the FOTRs  
and therefore $P(n|m)$ becomes purely non-perturbative. 
As an example for non-perturbative profile let 
us consider the lower plot of Fig.4, 
corresponding to $\delta x = 0.2123$.
We see that there is poor resemblance 
between $P(r)$ and $P_{\tbox{prt}}(r)$. 
The LDOS profile $P(r)$ no longer contains a predominant FOTRs.  
This claim can be quantified using the analysis 
in Fig.5. The lower subfigure there 
displays the FOTR contribution to the dispersion. 
For $\delta x > \delta x_{\tbox{prt}}$ 
the dispersion is no longer determined 
by the FOTR contribution.

The complete disappearance of FOTRs is guaranteed 
only for $\delta x \gg \delta x_{\tbox{prt}}$. 
Evidently, for $\delta x \gg \delta x_{\tbox{prt}}$ 
the FOTRs must disappear, because $P(r)$ goes on expanding, 
while $P_{\tbox{prt}}(r)$ saturates. 
This is not captured by our numerics since 
for $\hbar=0.015$ we cannot satisfy the strong inequality 
$\delta x \gg \delta x_{\tbox{prt}}$, and have 
a classically small $\delta x$ at the same time.

\section{The semiclassical regime}

Looking back at the lower plot of Fig.4, 
we see that detailed QCC with the classical profile 
(represented by heavy dashed line) starts to develop.  
The right far tail contains a component where $P(r)$ 
and $P_{\tbox{cl}}(r)$ are indistinguishable. 
This detailed QCC obviously does not hold for the RMT profile. 

Being in the non-perturbative regime does not imply 
detailed QCC \cite{crs,frc,vrn}. Detailed QCC means 
that $P(r)$ can be approximated by $P_{\tbox{cl}}(r)$. 
Having $\delta x \gg \delta x_{\tbox{prt}}$ is a necessary 
rather than sufficient condition for detailed QCC.

A sufficient condition for detailed QCC is 
$\delta x \gg \delta x_{\tbox{SC}}$. 
The parametric scale $\delta x_{\tbox{SC}}$  
is defined in \cite{frc}, and for our system we can 
obtain the (theoretical) rough estimate 
$\delta x_{\tbox{SC}}\approx 4*\hbar^{2/3}$. 

In our numerical study we could not make 
$\hbar$ small enough such that  
$\delta x_{\tbox{SC}} \ll \delta x_c^{\tbox{cl}}$.  
Therefore, the lower profile in Fig.4 
is neither reasonably approximated by 
$P_{\tbox{prt}}(r)$, nor by $P_{\tbox{cl}}(r)$. 
However, we have verified (by comparing the $\hbar=0.03$
output to the $\hbar=0.015$ output) that 
detailed QCC between $P(r)$ and $P_{\tbox{cl}}(r)$
is easily improved by making $\hbar$ smaller. 
Comparing $P(r)$ to $P_{\tbox{cl}}(r)$ on the one 
hand, and $P_{\tbox{rmt}}(r)$ to $P_{\tbox{cl}}(r)$ 
on the other hand, leaves no doubt regarding the 
manifestation of underlying classical structures.

Using a phase-space picture \cite{frc,vrn} it is evident 
that larger $\delta x$ leads to better QCC. 
The WBRM model does not have a classical limit, 
and one finds a quite different scenario \cite{casati}. 
For large enough $\delta x$  the eigenstates of 
Eq.(\ref{e1}) become Anderson localized. 
This localization shows up in the ASOE
{\em provided} the eigenstates are properly centered 
prior to averaging. In the (non-averaged) LDOS 
localization manifests itself as sparsity, 
and therefore the various moments of the LDOS profile 
are not affected. This latter remark should be kept 
in mind while reading the next section.

\ \\ \ \\
\epsfysize=2.6in
\epsffile{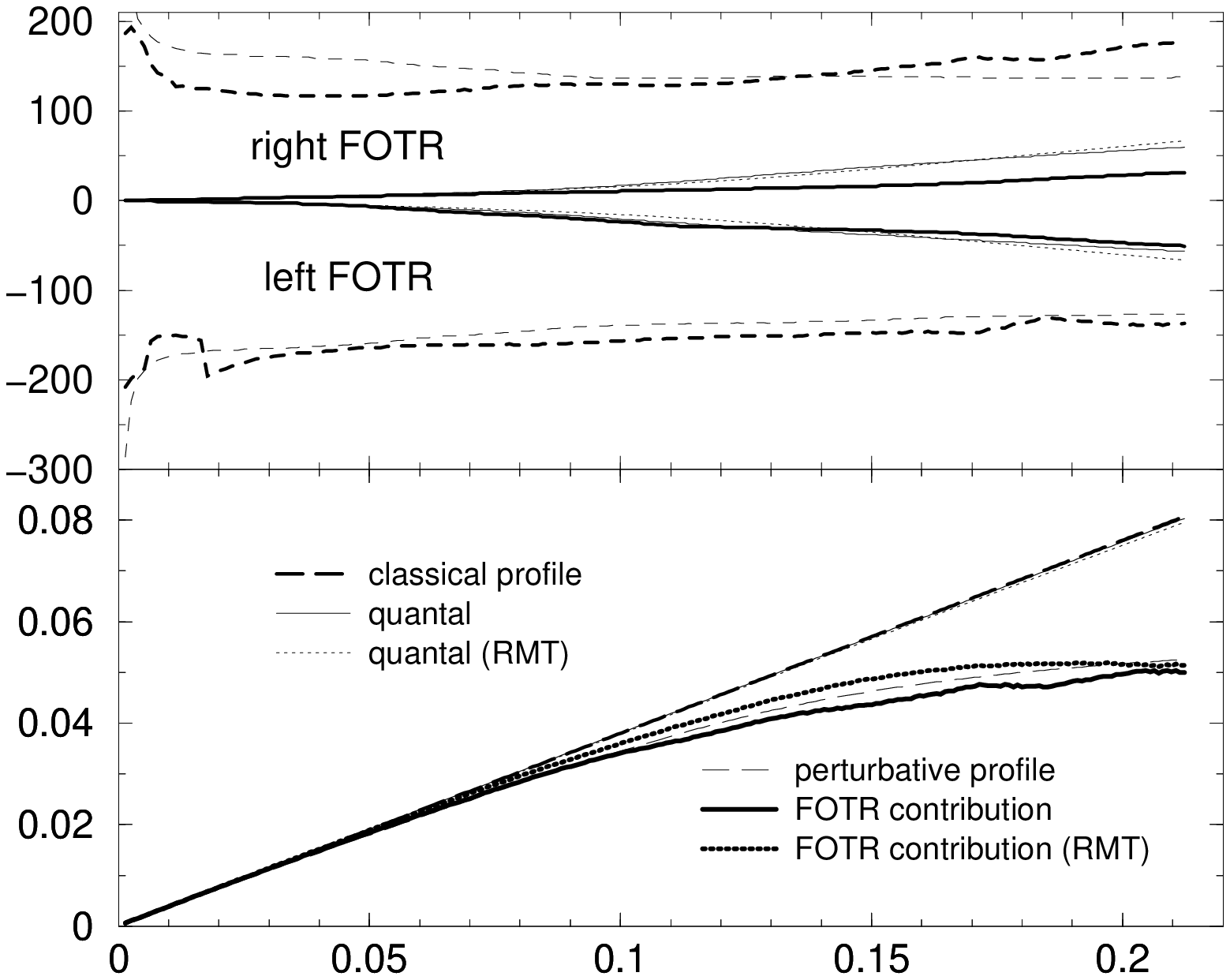} \\
\noindent
{\footnotesize {\bf FIG.5}
The results of region analysis. 
The common horizontal axis is $\delta x$. 
The {\em upper subfigure} presents the $r$ boundaries  
as a function of $\delta x$.  
The dotted lines $\pm b_0$ define the core region ($|r|<b_0$). 
The solid lines define the $r$ region in which $50\%$ 
of the probability is concentrate. 
The dashed lines are $b_1\mbox{[left]}$ and $b_1\mbox{[right]}$. 
The FOTRs are the regions where $b_0<|r|<b_1$. 
The light solid lines and the light dashed lines 
are for the effective WBRM model.
The {\em lower subfigure} displays the 
dependence of $\delta E_{\tbox{cl}}$ and $\delta E_{\tbox{qm}}$ 
and $\delta E_{\tbox{prt}}$ on $\delta x$. 
The quantal and the classical results 
are almost indistinguishable, whereas $\delta E_{\tbox{prt}}$ 
approaches saturation. 
The contribution of the FOTRs to $\delta E_{\tbox{qm}}$ 
is also displayed. } \\

\section{Restricted QCC}

It is important to distinguish between detailed QCC 
and restricted QCC. Let us denote the dispersion of 
the quantal LDOS profile by $\delta E_{\tbox{qm}}$, 
The corresponding classical quantity is given 
by Eq.(\ref{e_6}). The two types of QCC are defined 
as follows: 
\begin{itemize}
\item
Detailed QCC means \ \ \ $P(r) \approx P_{\tbox{cl}}(r)$  
\item
Restricted QCC means \ \ $\delta E_{\tbox{qm}} \approx \delta E_{\tbox{cl}}$
\end{itemize} 
Obviously restricted QCC is a trivial consequence 
of detailed QCC, but the converse is not true. 
It turns out that restricted QCC is much more 
robust than detailed QCC.  In Fig.5 we see that 
the dispersion $\delta E_{\tbox{qm}}$ of either 
$P(r)$ or $P_{\tbox{RMT}}(r)$ is almost indistinguishable
from $\delta E_{\tbox{cl}}$.  This is quite remarkable 
becuase the corresponding LDOS profiles 
(quantal versus classical) are very different!

It is important to realize that restricted QCC 
is implied by first order perturbation theory. 
If we use Eq.(\ref{e5}) and take into accound 
the FOTR dominace which is implied by 
$\delta x \ll \delta x_{\tbox{prt}}$, then we get simply
\begin{eqnarray} \label{e_pr}
\delta E_{\tbox{qm}}  =   
\sum_n P(n|m) \ (E_n{-}E_m)^2  =
\delta x^2 \sum_n'|\mbf{B}_{nm}|^2 
\end{eqnarray}
where prime indicates omission of the $n=m$ term.   
Using Eq.(\ref{e_5}) one realizes that this 
result is in complete agreement with Eq.(\ref{e_6}).
In contrast to that higher moments of the perturbative 
profile are vanishingly small compared with the corresponding 
classical result. The latter fact is just a reflection 
of the absence of detailed QCC.

One may wonder what happens with Eq.(\ref{e_pr})   
if we try to do a better work, taking into account the core width, 
as well as higher order far-tails contributions. 
One may think that Eq.(\ref{e_pr}) is only 
the lowest order approximation, which would imply 
that restricted QCC should become worse as $\delta x$ grows. 
However, the latter speculation turns out to be wrong.

We already saw that restricted QCC is implied 
on the one hand (for small $\delta x$) 
by first-order perturbation theory, and on the other 
hand (for large $\delta x$) by detailed QCC. 
Now we would like to argue that 
restricted QCC holds in general. 
It simply follows from the observation 
that $\delta E_{\tbox{qm}}$ is determined 
just by the band profile.
The prove is very simple \cite{casati}. 
The variance of $P(n|m)$ is determined 
by the first two moments of the Hamiltonian 
in the unperturbed basis. Namely
\begin{eqnarray} \nonumber 
\delta E_{\tbox{qm}}^2  =  
\langle m | {\cal H}^2 |m \rangle - \langle m| {\cal H} |m \rangle^2 
\\ \nonumber    
= \delta x^2
(\langle m | \mbf{B}^2 |m \rangle - \langle m| \mbf{B} |m \rangle^2)
\end{eqnarray}
Thus, we get the same result as in first order 
perturbation theory without invoking any special 
assumptions regarding the nature of the profile. 
Having  $\delta E_{\tbox{qm}}$ that is determined 
only by the band profile, is the reason for detailed QCC, 
and is also the reason why restricted QCC is not sensitive 
to the RMT assumption.   

\ \\


We thank Felix Izrailev for suggesting to study the 
model (\ref{e2}). We also thank ITAMP for their support. 



\end{multicols}

\begin{thebibliography}{99}

\bibitem[\dag]{note1}
To be more specific, one should be aware that there 
is an hierarchy of challenges where the applicability 
of the RMT approach should be tested. Namely: 
The study of spectral statistics; 
The study of eigenstates; 
The study of quantum dynamics. 
In a previous work \cite{wbr} we have argued that  
the RMT approach does not generally apply to the 
study of wavepacket dynamics, since it leads to 
a contradiction with the QCC principle.     
On the other hand, it is well known that spectral statistics 
is much more robust. In most of the RMT literature 
(including the later works by Wigner himself), 
it is assumed that for the purpose of 
`quantum chaos' studies one can consider 
full (rather than banded) matrices, and   
the first term in Eq.(\ref{e1}) is generally neglected. 
In spite of these enormous simplifications, 
it turns out that the so-called Gaussian invariant ensembles
(GOE,GUE) provide a valid description of some major 
spectral properties. 


\bibitem[\ddag]{note2}
The determination of $b_1$ has been done using the following 
numerical procedure. We define relative error function 
$\mbox{RE}(r) = (P-P_{\tbox{prt}})/(P+P_{\tbox{prt}})$ 
and then cumulative error 
function $\mbox{CRE}(r) = |\sum_0^r \mbox{RE}(r')|$.   
Note that by this definition `positive' relative error can 
be compensated by 'negative' relative error. 
As we go away from $r=0$, the function $\mbox{CRE}(r)$ fluctuates, 
and later shoots up. The regime $|r|<b_1$ 
has been determined  by the condition 
$\mbox{CRE}(r)<\mbox{Threshold}$. 
The threshold has been determined using adaptively procedure.  


\bibitem{mario}
M. Feingold and A. Peres, Phys. Rev. A {\bf 34} 591, (1986).
M. Feingold, D. Leitner, M. Wilkinson, Phys. Rev. Lett. {\bf 66}, 986 (1991). 
M. Wilkinson, M. Feingold, D. Leitner, J. Phys. A {\bf 24}, 175 (1991). 
M. Feingold, A. Gioletta, F. M. Izrailev, L. Molinari, Phys. Rev. Lett. 
{\bf 70}, 2936 (1993).

\bibitem{wigner}
E. Wigner, Ann. Math {\bf 62} 548 (1955); {\bf 65} 203 (1957).

\bibitem{casati}
G. Casati, B.V. Chirikov, I. Guarneri, F.M. Izrailev, 
Phys. Rev. E {\bf 48}, R1613 (1993);
\ Phys. Lett. A {\bf 223}, 430 (1996).
V.V. Flambaum, A.A. Gribakina, G.F. Gribakin and M.G. Kozlov, 
Phys. Rev. A {\bf 50} 267 (1994).

\bibitem{fyodo}
Y.V. Fyodorov, O.A. Chubykalo, F.M. Izrailev and G. Casati, 
Phys. Rev. Lett. {\bf 76}, 1603 (1996).

\bibitem{frc}
D. Cohen, Ann. Phys. {\bf 283}, 175-231 (2000).
 
\bibitem{vrn} D. Cohen, lecture notes, 
{\em in} Proceedings of the International 
School of Physics `Enrico Fermi' Course CXLIII 
``New Directions in Quantum Chaos'', 
Edited by G. Casati, I. Guarneri and U. Smilansky, 
IOS Press, Amsterdam, 2000. 

\bibitem{wls} 
D. Cohen and E.J. Heller,  
Phys. Rev. Lett. {\bf 84}, 2841 (2000). 

\bibitem{prm} 
In a later work regarding "Parametric Evolution for a Deformed Cavity",    
(D. Cohen, A. Barnett and E.J. Heller, nlin.CD/0008040), 
further study of the various parametric regimes is introduced. 
In our present paper the emphasis is on generic features of 
parametric evolution, while in nlin.CD/0008040 the emphasize is 
on non-generic features which have been discussed in \cite{wls}.     
 
\bibitem{felix2}
F. Borgonovi, I. Guarneri and F.M. Izrailev, 
Phys. Rev. E {\bf 57}, 5291 (1998). 
L. Benet, F.M. Izrailev, T.H. Seligman and A. Suarez-Moreno, 
chao-dyn/9912035. 

\bibitem{crs}
D. Cohen, Phys. Rev. Lett. {\bf 82}, 4951 (1999) 

\bibitem{wbr} 
D. Cohen, F.M. Izrailev and T. Kottos,  
Phys. Rev. Lett. {\bf 84} 2052 (2000). 

\end{thebibliography}
\end{document}